
%
%
%
\def\unredoffs{} \def\redoffs{\voffset=-.31truein\hoffset=-.59truein}
\def\speclscape{\special{ps: landscape}}
%
%
%
%
\newbox\leftpage \newdimen\fullhsize \newdimen\hstitle \newdimen\hsbody
\tolerance=1000\hfuzz=2pt
\catcode`\@=11 
\def\bigans{b }
\def\answ{b }
\ifx\answ\bigans\message{(This will come out unreduced.}
\magnification=1200\unredoffs\baselineskip=16pt plus 2pt minus 1pt
\hsbody=\hsize \hstitle=\hsize 
\else\message{(This will be reduced.} \let\l@r=L
\magnification=1000\baselineskip=16pt plus 2pt minus 1pt \vsize=7truein
\redoffs \hstitle=8truein\hsbody=4.75truein\fullhsize=10truein\hsize=\hsbody
\output={\ifnum\pageno=0 
  \shipout\vbox{\speclscape{\hsize\fullhsize\makeheadline}
    \hbox to \fullhsize{\hfill\pagebody\hfill}}\advancepageno
  \else
  \almostshipout{\leftline{\vbox{\pagebody\makefootline}}}\advancepageno
  \fi}
\def\almostshipout#1{\if L\l@r \count1=1 \message{[\the\count0.\the\count1]}
      \global\setbox\leftpage=#1 \global\let\l@r=R
 \else \count1=2
  \shipout\vbox{\speclscape{\hsize\fullhsize\makeheadline}
      \hbox to\fullhsize{\box\leftpage\hfil#1}}  \global\let\l@r=L\fi}
\fi
%
\newcount\yearltd\yearltd=\year\advance\yearltd by -1900

\def\Title#1#2{\nopagenumbers\abstractfont\hsize=\hstitle\rightline{#1}%
\vskip 1in\centerline{\titlefont #2}\abstractfont\vskip .5in\pageno=0}
\def\Date#1{\vfill\leftline{#1}\tenpoint\supereject\global\hsize=\hsbody%
\footline={\hss\tenrm\folio\hss}}
%

\def\draftmode{\message{ DRAFTMODE }\def\draftdate{{\rm preliminary draft:
\number\month/\number\day/\number\yearltd\ \ \hourmin}}%
\headline={\hfil\draftdate}\writelabels\baselineskip=20pt plus 2pt minus 2pt
 {\count255=\time\divide\count255 by 60 \xdef\hourmin{\number\count255}
  \multiply\count255 by-60\advance\count255 by\time
  \xdef\hourmin{\hourmin:\ifnum\count255<10 0\fi\the\count255}}}
\def\nolabels{\def\wrlabeL##1{}\def\eqlabeL##1{}\def\reflabeL##1{}}
\def\writelabels{\def\wrlabeL##1{\leavevmode\vadjust{\rlap{\smash%
{\line{{\escapechar=` \hfill\rlap{\sevenrm\hskip.03in\string##1}}}}}}}%
\def\eqlabeL##1{{\escapechar-1\rlap{\sevenrm\hskip.05in\string##1}}}%
\def\reflabeL##1{\noexpand\llap{\noexpand\sevenrm\string\string\string##1}}}
\nolabels
%
\global\newcount\secno \global\secno=0
\global\newcount\meqno \global\meqno=1
\def\newsec#1{\global\advance\secno by1\message{(\the\secno. #1)}
\global\subsecno=0\eqnres@t\noindent{\bf\the\secno. #1}
\writetoca{{\secsym} {#1}}\par\nobreak\medskip\nobreak}
\def\eqnres@t{\xdef\secsym{\the\secno.}\global\meqno=1\bigbreak\bigskip}
\def\sequentialequations{\def\eqnres@t{\bigbreak}}\xdef\secsym{}
\global\newcount\subsecno \global\subsecno=0
\def\subsec#1{\global\advance\subsecno by1\message{(\secsym\the\subsecno.
#1)}
\ifnum\lastpenalty>9000\else\bigbreak\fi
\noindent{\it\secsym\the\subsecno. #1}\writetoca{\string\quad
{\secsym\the\subsecno.} {#1}}\par\nobreak\medskip\nobreak}
\def\appendix#1#2{\global\meqno=1\global\subsecno=0\xdef\secsym{\hbox{#1.}}
\bigbreak\bigskip\noindent{\bf Appendix #1. #2}\message{(#1. #2)}
\writetoca{Appendix {#1.} {#2}}\par\nobreak\medskip\nobreak}
%
%
\def\eqnn#1{\xdef #1{(\secsym\the\meqno)}\writedef{#1\leftbracket#1}%
\global\advance\meqno by1\wrlabeL#1}
\def\eqna#1{\xdef #1##1{\hbox{$(\secsym\the\meqno##1)$}}
\writedef{#1\numbersign1\leftbracket#1{\numbersign1}}%
\global\advance\meqno by1\wrlabeL{#1$\{\}$}}
\def\eqn#1#2{\xdef #1{(\secsym\the\meqno)}\writedef{#1\leftbracket#1}%
\global\advance\meqno by1$$#2\eqno#1\eqlabeL#1$$}
%
\newskip\footskip\footskip14pt plus 1pt minus 1pt 
\def\footnotefont{\ninepoint}\def\f@t#1{\footnotefont #1\@foot}
\def\f@@t{\baselineskip\footskip\bgroup\footnotefont\aftergroup\@foot\let\next}
\setbox\strutbox=\hbox{\vrule height9.5pt depth4.5pt width0pt}
\global\newcount\ftno \global\ftno=0
\def\foot{\global\advance\ftno by1\footnote{$^{\the\ftno}$}}
%
\newwrite\ftfile
\def\footend{\def\foot{\global\advance\ftno by1\chardef\wfile=\ftfile
$^{\the\ftno}$\ifnum\ftno=1\immediate\openout\ftfile=foots.tmp\fi%
\immediate\write\ftfile{\noexpand\smallskip%
\noexpand\item{f\the\ftno:\ }\pctsign}\findarg}%
\def\footatend{\vfill\eject\immediate\closeout\ftfile{\parindent=20pt
\centerline{\bf Footnotes}\nobreak\bigskip\input foots.tmp }}}
\def\footatend{}
%
%
\global\newcount\refno \global\refno=1
\newwrite\rfile
\def\ref{[\the\refno]\nref}
\def\nref#1{\xdef#1{[\the\refno]}\writedef{#1\leftbracket#1}%
\ifnum\refno=1\immediate\openout\rfile=refs.tmp\fi
\global\advance\refno by1\chardef\wfile=\rfile\immediate
\write\rfile{\noexpand\item{#1\ }\reflabeL{#1\hskip.31in}\pctsign}\findarg}
\def\findarg#1#{\begingroup\obeylines\newlinechar=`\^^M\pass@rg}
{\obeylines\gdef\pass@rg#1{\writ@line\relax #1^^M\hbox{}^^M}%
\gdef\writ@line#1^^M{\expandafter\toks0\expandafter{\striprel@x #1}%
\edef\next{\the\toks0}\ifx\next\em@rk\let\next=\endgroup\else\ifx\next\empty%
\else\immediate\write\wfile{\the\toks0}\fi\let\next=\writ@line\fi\next\relax}}
\def\striprel@x#1{} \def\em@rk{\hbox{}}
\def\lref{\begingroup\obeylines\lr@f}
\def\lr@f#1#2{\gdef#1{\ref#1{#2}}\endgroup\unskip}
\def\semi{;\hfil\break}
\def\addref#1{\immediate\write\rfile{\noexpand\item{}#1}} 
\def\startrefs#1{\immediate\openout\rfile=refs.tmp\refno=#1}
\def\xref{\expandafter\xr@f}\def\xr@f[#1]{#1}
\def\refs#1{\count255=1[\r@fs #1{\hbox{}}]}
\def\r@fs#1{\ifx\und@fined#1\message{reflabel \string#1 is undefined.}%
\nref#1{need to supply reference \string#1.}\fi%
\vphantom{\hphantom{#1}}\edef\next{#1}\ifx\next\em@rk\def\next{}%
\else\ifx\next#1\ifodd\count255\relax\xref#1\count255=0\fi%
\else#1\count255=1\fi\let\next=\r@fs\fi\next}
%

%
\newwrite\ffile\global\newcount\figno \global\figno=1
\def\fig{fig.~\the\figno\nfig}
\def\nfig#1{\xdef#1{fig.~\the\figno}%
\writedef{#1\leftbracket fig.\noexpand~\the\figno}%
\ifnum\figno=1\immediate\openout\ffile=figs.tmp\fi\chardef\wfile=\ffile%
\immediate\write\ffile{\noexpand\medskip\noexpand\item{Fig.\ \the\figno. }
\reflabeL{#1\hskip.55in}\pctsign}\global\advance\figno by1\findarg}
\def\xfig{\expandafter\xf@g}\def\xf@g fig.\penalty\@M\ {}
\def\figs#1{figs.~\f@gs #1{\hbox{}}}
\def\f@gs#1{\edef\next{#1}\ifx\next\em@rk\def\next{}\else
\ifx\next#1\xfig #1\else#1\fi\let\next=\f@gs\fi\next}
\newwrite\lfile
{\escapechar-1\xdef\pctsign{\string\%}\xdef\leftbracket{\string\{}
\xdef\rightbracket{\string\}}\xdef\numbersign{\string\#}}

\def\writestop{\def\writestoppt{\immediate\write\lfile{\string\pageno%
\the\pageno\string\startrefs\leftbracket\the\refno\rightbracket%
\string\def\string\secsym\leftbracket\secsym\rightbracket%
\string\secno\the\secno\string\meqno\the\meqno}\immediate\closeout\lfile}}
\def\writestoppt{}\def\writedef#1{}
\def\seclab#1{\xdef #1{\the\secno}\writedef{#1\leftbracket#1}\wrlabeL{#1=#1}}
\def\subseclab#1{\xdef #1{\secsym\the\subsecno}%
\writedef{#1\leftbracket#1}\wrlabeL{#1=#1}}
\newwrite\tfile \def\writetoca#1{}
\def\leaderfill{\leaders\hbox to 1em{\hss.\hss}\hfill}
\def\writetoc{\immediate\openout\tfile=toc.tmp
   \def\writetoca##1{{\edef\next{\write\tfile{\noindent ##1
   \string\leaderfill {\noexpand\number\pageno} \par}}\next}}}

%
\catcode`\@=12 
%
\edef\tfontsize{\ifx\answ\bigans scaled\magstep3\else scaled\magstep4\fi}
\font\titlerm=cmr10 \tfontsize \font\titlerms=cmr7 \tfontsize
\font\titlermss=cmr5 \tfontsize \font\titlei=cmmi10 \tfontsize
\font\titleis=cmmi7 \tfontsize \font\titleiss=cmmi5 \tfontsize
\font\titlesy=cmsy10 \tfontsize \font\titlesys=cmsy7 \tfontsize
\font\titlesyss=cmsy5 \tfontsize \font\titleit=cmti10 \tfontsize
\skewchar\titlei='177 \skewchar\titleis='177 \skewchar\titleiss='177
\skewchar\titlesy='60 \skewchar\titlesys='60 \skewchar\titlesyss='60
\def\titlefont{\def\rm{\fam0\titlerm}
\textfont0=\titlerm \scriptfont0=\titlerms \scriptscriptfont0=\titlermss
\textfont1=\titlei \scriptfont1=\titleis \scriptscriptfont1=\titleiss
\textfont2=\titlesy \scriptfont2=\titlesys \scriptscriptfont2=\titlesyss
\textfont\itfam=\titleit \def\it{\fam\itfam\titleit}\rm}
 \ifx\answ\bigans\else scaled\magstep1\fi
\ifx\answ\bigans\def\abstractfont{\tenpoint}\else
\font\abssl=cmsl10 scaled \magstep1
\font\absrm=cmr10 scaled\magstep1 \font\absrms=cmr7 scaled\magstep1
\font\absrmss=cmr5 scaled\magstep1 \font\absi=cmmi10 scaled\magstep1
\font\absis=cmmi7 scaled\magstep1 \font\absiss=cmmi5 scaled\magstep1
\font\abssy=cmsy10 scaled\magstep1 \font\abssys=cmsy7 scaled\magstep1
\font\abssyss=cmsy5 scaled\magstep1 \font\absbf=cmbx10 scaled\magstep1
\skewchar\absi='177 \skewchar\absis='177 \skewchar\absiss='177
\skewchar\abssy='60 \skewchar\abssys='60 \skewchar\abssyss='60
\def\abstractfont{\def\rm{\fam0\absrm}
\textfont0=\absrm \scriptfont0=\absrms \scriptscriptfont0=\absrmss
\textfont1=\absi \scriptfont1=\absis \scriptscriptfont1=\absiss
\textfont2=\abssy \scriptfont2=\abssys \scriptscriptfont2=\abssyss
\textfont\itfam=\bigit \def\it{\fam\itfam\bigit}\def\footnotefont{\tenpoint}%
\textfont\slfam=\abssl \def\sl{\fam\slfam\abssl}%
\textfont\bffam=\absbf \def\bf{\fam\bffam\absbf}\rm}\fi
\def\tenpoint{\def\rm{\fam0\tenrm}
\textfont0=\tenrm \scriptfont0=\sevenrm \scriptscriptfont0=\fiverm
\textfont1=\teni  \scriptfont1=\seveni  \scriptscriptfont1=\fivei
\textfont2=\tensy \scriptfont2=\sevensy \scriptscriptfont2=\fivesy
\textfont\itfam=\tenit
\def\it{\fam\itfam\tenit}\def\footnotefont{\ninepoint}%
\textfont\bffam=\tenbf \def\bf{\fam\bffam\tenbf}\def\sl{\fam\slfam\tensl}\rm}
\font\ninerm=cmr9 \font\sixrm=cmr6 \font\ninei=cmmi9 \font\sixi=cmmi6
\font\ninesy=cmsy9 \font\sixsy=cmsy6 \font\ninebf=cmbx9
\font\nineit=cmti9 \font\ninesl=cmsl9 \skewchar\ninei='177
\skewchar\sixi='177 \skewchar\ninesy='60 \skewchar\sixsy='60
\def\ninepoint{\def\rm{\fam0\ninerm}
\textfont0=\ninerm \scriptfont0=\sixrm \scriptscriptfont0=\fiverm
\textfont1=\ninei \scriptfont1=\sixi \scriptscriptfont1=\fivei
\textfont2=\ninesy \scriptfont2=\sixsy \scriptscriptfont2=\fivesy
\textfont\itfam=\ninei \def\it{\fam\itfam\nineit}\def\sl{\fam\slfam\ninesl}%
\textfont\bffam=\ninebf \def\bf{\fam\bffam\ninebf}\rm}
%
%

\hyphenation{anom-aly anom-alies coun-ter-term coun-ter-terms}
\def\inv{^{\raise.15ex\hbox{${\scriptscriptstyle -}$}\kern-.05em 1}}

\def\Dsl{\,\raise.15ex\hbox{/}\mkern-13.5mu D} 
\def\dsl{\raise.15ex\hbox{/}\kern-.57em\partial}

\font\bigit=cmti10 scaled \magstep1
\def\lspace{\ifx\answ\bigans{}\else\qquad\fi}
\def\lbspace{\ifx\answ\bigans{}\else\hskip-.2in\fi} 
\def\boxeqn#1{\vcenter{\vbox{\hrule\hbox{\vrule\kern3pt\vbox{\kern3pt
           \hbox{${\displaystyle #1}$}\kern3pt}\kern3pt\vrule}\hrule}}}
\def\mbox#1#2{\vcenter{\hrule \hbox{\vrule height#2in
               \kern#1in \vrule} \hrule}}  
%

\def\e#1{{\rm e}^{^{\textstyle#1}}}

\def\om#1#2{\omega^{#1}{}_{#2}}

\def\darr#1{\raise1.5ex\hbox{$\leftrightarrow$}\mkern-16.5mu #1}

\def\roughly#1{\raise.3ex\hbox{$#1$\kern-.75em\lower1ex\hbox{$\sim$}}}



\def\IB{\relax\hbox{$\inbar\kern-.3em{\rm B}$}}
\def\IC{\relax\hbox{$\inbar\kern-.3em{\rm C}$}}
\def\ID{\relax\hbox{$\inbar\kern-.3em{\rm D}$}}
\def\IE{\relax\hbox{$\inbar\kern-.3em{\rm E}$}}
\def\IF{\relax\hbox{$\inbar\kern-.3em{\rm F}$}}
\def\IG{\relax\hbox{$\inbar\kern-.3em{\rm G}$}}
\def\IGa{\relax\hbox{${\rm I}\kern-.18em\Gamma$}}
\def\IH{\relax{\rm I\kern-.18em H}}
\def\IK{\relax{\rm I\kern-.18em K}}
\def\II{\relax{\rm I\kern-.18em I}}
\def\IL{\relax{\rm I\kern-.18em L}}
\def\IP{\relax{\rm I\kern-.18em P}}
\def\IR{\relax{\rm I\kern-.18em R}}
\def\IZ{\relax\ifmmode\mathchoice {\hbox{\cmss Z\kern-.4em Z}}{\hbox{\cmss
Z\kern-.4em Z}} {\lower.9pt\hbox{\cmsss Z\kern-.4em Z}}
{\lower1.2pt\hbox{\cmsss Z\kern-.4em Z}}\else{\cmss Z\kern-.4em Z}\fi}

\def\IB{\relax{\rm I\kern-.18em B}}
\def\IC{{\relax\hbox{$\inbar\kern-.3em{\rm C}$}}}
\def\ID{\relax{\rm I\kern-.18em D}}
\def\IE{\relax{\rm I\kern-.18em E}}
\def\IF{\relax{\rm I\kern-.18em F}}


\def\CW {{\cal W}}

\def\p{\partial}




\def\s{\lies}


\def\demi{{1\over 2}}


\def\a{\alpha}

  \def\G{\Gamma}
\def\d{\delta}

\def\l{\lambda} \def\L{\Lambda}

\def\e{\epsilon}

\def\|{\Big|}
\def\({\Big(}   \def\){\Big)}
\def\[{\Big[}   \def\]{\Big]}



\def\paper#1#2#3#4{#1, {\sl #2}, #3 {\tt #4}}

\def\hh{hep-th/}


\def\PLB#1#2#3{Phys. Lett.~{\bf B#1} (#2) #3}
\def\NPB#1#2#3{Nucl. Phys.~{\bf B#1} (#2) #3}
\def\PRL#1#2#3{Phys. Rev. Lett.~{\bf #1} (#2) #3}
\def\CMP#1#2#3{Comm. Math. Phys.~{\bf #1} (#2) #3}
\def\PRD#1#2#3{Phys. Rev.~{\bf D#1} (#2) #3}
\def\MPL#1#2#3{Mod. Phys. Lett.~{\bf #1} (#2) #3}
\def\IJMP#1#2#3{Int. Jour. Mod. Phys.~{\bf #1} (#2) #3}


\def\unlockat{\catcode`\@=11}
\def\lockat{\catcode`\@=12}

\unlockat


\def\newsec#1{\global\advance\secno by1\message{(\the\secno. #1)}
\global\subsecno=0\global\subsubsecno=0\eqnres@t\noindent {\bf\the\secno. #1}
\writetoca{{\secsym} {#1}}\par\nobreak\medskip\nobreak}
\global\newcount\subsecno \global\subsecno=0
\def\subsec#1{\global\advance\subsecno by1\message{(\secsym\the\subsecno.
#1)}
\ifnum\lastpenalty>9000\else\bigbreak\fi\global\subsubsecno=0
\noindent{\it\secsym\the\subsecno. #1}
\writetoca{\string\quad {\secsym\the\subsecno.} {#1}}
\par\nobreak\medskip\nobreak}
\global\newcount\subsubsecno \global\subsubsecno=0
\def\subsubsec#1{\global\advance\subsubsecno by1
\message{(\secsym\the\subsecno.\the\subsubsecno. #1)}
\ifnum\lastpenalty>9000\else\bigbreak\fi
\noindent\quad{\secsym\the\subsecno.\the\subsubsecno.}{#1}
\writetoca{\string\qquad{\secsym\the\subsecno.\the\subsubsecno.}{#1}}
\par\nobreak\medskip\nobreak}

\def\subsubseclab#1{\DefWarn#1\xdef #1{\noexpand\hyperref{}{subsubsection}%
{\secsym\the\subsecno.\the\subsubsecno}%
{\secsym\the\subsecno.\the\subsubsecno}}%
\writedef{#1\leftbracket#1}\wrlabeL{#1=#1}}
\lockat

\def\dbend{\lower3.5pt\hbox{\manual\char127}}


\def\boxit#1{\vbox{\hrule\hbox{\vrule\kern8pt
\vbox{\hbox{\kern8pt}\hbox{\vbox{#1}}\hbox{\kern8pt}}
\kern8pt\vrule}\hrule}}

\def\mathboxit#1{\vbox{\hrule\hbox{\vrule\kern8pt\vbox{\kern8pt
\hbox{$\displaystyle #1$}\kern8pt}\kern8pt\vrule}\hrule}}


\def\inbar{\,\vrule height1.5ex width.4pt depth0pt}

\font\cmss=cmss10 \font\cmsss=cmss10 at 7pt


\lref\simons{ J. Cheeger and J. Simons, {\it Differential Characters and
Geometric Invariants},  Stony Brook Preprint, (1973), unpublished.}

\lref\cargese{ L.~Baulieu, {\it Algebraic quantization of gauge theories},
Perspectives in fields and particles, Plenum Press, eds. Basdevant-Levy,
Cargese Lectures 1983}

\lref\antifields{ L. Baulieu, M. Bellon, S. Ouvry, C.Wallet, Phys.Letters
B252 (1990) 387; M.  Bocchichio, Phys. Lett. B187 (1987) 322;  Phys. Lett. B
192 (1987) 31; R.  Thorn    Nucl. Phys.   B257 (1987) 61. }

\lref\thompson{ George Thompson,  Annals Phys. 205 (1991) 130; J.M.F.
Labastida, M. Pernici, Phys. Lett. 212B  (1988) 56; D. Birmingham, M.Blau,
M. Rakowski and G.Thompson, Phys. Rept. 209 (1991) 129.}

\lref\tonin{ Tonin}

\lref\wittensix{ E.  Witten, {\it New  Gauge  Theories In Six Dimensions},
\hh{9710065}. }

\lref\orlando{ O. Alvarez, L. A. Ferreira and J. Sanchez Guillen, {\it  A New
Approach to Integrable Theories in any Dimension}, hep-th/9710147.}

\lref\wittentopo{ E.  Witten,  {\it  Topological Quantum Field Theory},
\hh9403195, Commun.  Math. Phys.  {117} (1988)353.  }

\lref\wittentwist{ E.  Witten, {\it Supersymmetric Yang--Mills theory on a
four-manifold}, J.  Math.  Phys.  {35} (1994) 5101.}

\lref\west{ L.~Baulieu, P.~West, {\it Six Dimensional TQFTs and  Self-dual
Two-Forms,} Phys.Lett. B {\bf 436 } (1998) 97, /hep-th/9805200}

\lref\bv{ I.A. Batalin and V.A. Vilkowisky,    Phys. Rev.   D28  (1983)
2567\semi M. Henneaux,  Phys. Rep.  126   (1985) 1\semi M. Henneaux and C.
Teitelboim, {\it Quantization of Gauge Systems}
  Princeton University Press,  Princeton (1992).}

\lref\kyoto{ L. Baulieu, E. Bergschoeff and E. Sezgin, Nucl. Phys.
B307(1988)348\semi L. Baulieu,   {\it Field Antifield Duality, p-Form Gauge
Fields
   and Topological Quantum Field Theories}, hep-th/9512026,
   Nucl. Phys. B478 (1996) 431.  }

\lref\sourlas{ G. Parisi and N. Sourlas, {\it Random Magnetic Fields,
Supersymmetry and Negative Dimensions}, Phys. Rev. Lett.  43 (1979) 744;
Nucl.  Phys.  B206 (1982) 321.  }

\lref\SalamSezgin{ A.  Salam  and  E.  Sezgin, {\it Supergravities in
diverse dimensions}, vol.  1, p. 119\semi P.  Howe, G.  Sierra and P.
Townsend, Nucl Phys B221 (1983) 331.}

\lref\nekrasov{ A. Losev, G. Moore, N. Nekrasov, S. Shatashvili, {\it
Four-Dimensional Avatars of Two-Dimensional RCFT},  hep-th/9509151, Nucl.
Phys.  Proc.  Suppl.   46 (1996) 130\semi L.  Baulieu, A.  Losev,
N.~Nekrasov  {\it Chern-Simons and Twisted Supersymmetry in Higher
Dimensions},  hep-th/9707174, to appear in Nucl.  Phys.  B.  }

\lref\WitDonagi{R.~ Donagi, E.~ Witten, ``Supersymmetric Yang--Mills Theory
and Integrable Systems'', hep-th/9510101, Nucl. Phys.{\bf B}460 (1996)
299-334}
\lref\Witfeb{E.~ Witten, ``Supersymmetric Yang--Mills Theory On A
Four-Manifold,''  hep-th/9403195; J. Math. Phys. {\bf 35} (1994) 5101.}
\lref\Witgrav{E.~ Witten, ``Topological Gravity'', Phys.Lett.206B:601, 1988}
\lref\witaffl{I. ~ Affleck, J.A.~ Harvey and E.~ Witten,
        ``Instantons and (Super)Symmetry Breaking
        in $2+1$ Dimensions'', Nucl. Phys. {\bf B}206 (1982) 413}
\lref\wittabl{E.~ Witten,  ``On $S$-Duality in Abelian Gauge Theory,''
hep-th/9505186; Selecta Mathematica {\bf 1} (1995) 383}
\lref\wittgr{E.~ Witten, ``The Verlinde Algebra And The Cohomology Of The
Grassmannian'',  hep-th/9312104}
\lref\wittenwzw{E. Witten, ``Non abelian bosonization in two dimensions,''
Commun. Math. Phys. {\bf 92} (1984)455 }
\lref\witgrsm{E. Witten, ``Quantum field theory, grassmannians and algebraic
curves,'' Commun.Math.Phys.113:529,1988}
\lref\wittjones{E. Witten, ``Quantum field theory and the Jones
polynomial,'' Commun.  Math. Phys., 121 (1989) 351. }
\lref\witttft{E.~ Witten, ``Topological Quantum Field Theory", Commun. Math.
Phys. {\bf 117} (1988) 353.}
\lref\wittmon{E.~ Witten, ``Monopoles and Four-Manifolds'', hep-th/9411102}
\lref\Witdgt{ E.~ Witten, ``On Quantum gauge theories in two dimensions,''
Commun. Math. Phys. {\bf  141}  (1991) 153}
\lref\witrevis{E.~ Witten,
 ``Two-dimensional gauge theories revisited'', hep-th/9204083; J. Geom.
Phys. 9 (1992) 303-368}
\lref\Witgenus{E.~ Witten, ``Elliptic Genera and Quantum Field Theory'',
Comm. Math. Phys. 109(1987) 525. }
\lref\OldZT{E. Witten, ``New Issues in Manifolds of SU(3) Holonomy,'' {\it
Nucl. Phys.} {\bf B268} (1986) 79 \semi J. Distler and B. Greene, ``Aspects
of (2,0) String Compactifications,'' {\it Nucl. Phys.} {\bf B304} (1988) 1
\semi B. Greene, ``Superconformal Compactifications in Weighted Projective
Space,'' {\it Comm. Math. Phys.} {\bf 130} (1990) 335.}
\lref\bagger{E.~ Witten and J. Bagger, Phys. Lett. {\bf 115B}(1982) 202}
\lref\witcurrent{E.~ Witten,``Global Aspects of Current Algebra'',
Nucl.Phys.B223 (1983) 422\semi ``Current Algebra, Baryons and Quark
Confinement'', Nucl.Phys. B223 (1993) 433}
\lref\Wittreiman{S.B. Treiman, E. Witten, R. Jackiw, B. Zumino, ``Current
Algebra and Anomalies'', Singapore, Singapore: World Scientific ( 1985) }
\lref\Witgravanom{L. Alvarez-Gaume, E.~ Witten, ``Gravitational Anomalies'',
Nucl.Phys.B234:269,1984. }

\lref\nicolai{\paper {H.~Nicolai}{New Linear Systems for 2D Poincar\'e
Supergravities}{\NPB{414}{1994}{299},}{\hh 9309052}.}



\lref\bg{\paper {L.~Baulieu, B.~Grossman}{Monopoles and Topological Field
Theory}{\PLB{214}{1988}{223}.}{}}

\lref\seibergsix{\paper {N.~Seiberg}{Non-trivial Fixed Points of The
Renormalization Group in Six
 Dimensions}{\PLB{390}{1997}{169}}{\hh 9609161}\semi
\paper {O.J.~Ganor, D.R.~Morrison, N.~Seiberg}{
  Branes, Calabi-Yau Spaces, and Toroidal Compactification of the N=1
  Six-Dimensional $E_8$ Theory}{\NPB{487}{1997}{93}}{\hh 9610251}\semi
\paper {O.~Aharony, M.~Berkooz, N.~Seiberg}{Light-Cone
  Description of (2,0) Superconformal Theories in Six
  Dimensions}{Adv. Theor. Math. Phys. {\bf 2} (1998) 119}{\hh 9712117.}}

\lref\cs{\paper {L.~Baulieu}{Chern-Simons Three-Dimensional and
Yang--Mills-Higgs Two-Dimensional Systems as Four-Dimensional Topological
Quantum Field Theories}{\PLB{232}{1989}{473}.}{}}

\lref\beltrami{\paper {L.~Baulieu, M.~Bellon}{Beltrami Parametrization and
String Theory}{\PLB{196}{1987}{142}}{}\semi
\paper {L.~Baulieu, M.~Bellon, R.~Grimm}{Beltrami Parametrization For
Superstrings}{\PLB{198}{1987}{343}}{}\semi
\paper {R.~Grimm}{Left-Right Decomposition of Two-Dimensional Superspace
Geometry and Its BRS Structure}{Annals Phys. {\bf 200} (1990) 49.}{}}

\lref\bbg{\paper {L.~Baulieu, M.~Bellon, R.~Grimm}{Left-Right Asymmetric
Conformal Anomalies}{\PLB{228}{1989}{325}.}{}}

\lref\bonora{\paper {G.~Bonelli, L.~Bonora, F.~Nesti}{String Interactions
from Matrix String Theory}{\NPB{538}{1999}{100},}{\hh 9807232}\semi
\paper {G.~Bonelli, L.~Bonora, F.~Nesti, A.~Tomasiello}{Matrix String Theory
and its Moduli Space}{}{\hh 9901093.}}

\lref\corrigan{\paper {E.~Corrigan, C.~Devchand, D.B.~Fairlie,
J.~Nuyts}{First Order Equations for Gauge Fields in Spaces of Dimension
Greater Than Four}{\NPB{214}{452}{1983}.}{}}

\lref\acha{\paper {B.S.~Acharya, M.~O'Loughlin, B.~Spence}{Higher
Dimensional Analogues of Donaldson-Witten Theory}{\NPB{503}{1997}{657},}{\hh
9705138}\semi
\paper {B.S.~Acharya, J.M.~Figueroa-O'Farrill, M.~O'Loughlin,
B.~Spence}{Euclidean
  D-branes and Higher-Dimensional Gauge   Theory}{\NPB{514}{1998}{583},}{\hh
  9707118.}}

\lref\Witr{\paper{E.~Witten}{Introduction to Cohomological Field   Theories}
{Lectures at Workshop on Topological Methods in Physics (Trieste, Italy, Jun
11-25, 1990), \IJMP{A6}{1991}{2775}.}{}}

\lref\ohta{\paper {L.~Baulieu, N.~Ohta}{Worldsheets with Extended
Supersymmetry} {\PLB{391}{1997}{295},}{\hh 9609207}.}

\lref\gravity{\paper {L.~Baulieu}{Transmutation of Pure 2-D Supergravity
Into Topological 2-D Gravity and Other Conformal Theories}
{\PLB{288}{1992}{59},}{\hh 9206019.}}

\lref\wgravity{\paper {L.~Baulieu, M.~Bellon, R.~Grimm}{Some Remarks on  the
Gauging of the Virasoro and   $w_{1+\infty}$
Algebras}{\PLB{260}{1991}{63}.}{}}

\lref\fourd{\paper {E.~Witten}{Topological Quantum Field
Theory}{\CMP{117}{1988}{353}}{}\semi
\paper {L.~Baulieu, I.M.~Singer}{Topological Yang--Mills Symmetry}{Nucl.
Phys. Proc. Suppl. {\bf 15B} (1988) 12.}{}}

\lref\topo{\paper {L.~Baulieu}{On the Symmetries of Topological Quantum Field
Theories}{\IJMP{A10}{1995}{4483},}{\hh 9504015}\semi
\paper {R.~Dijkgraaf, G.~Moore}{Balanced Topological Field
Theories}{\CMP{185}{1997}{411},}{\hh 9608169.}}

\lref\wwgravity{\paper {I.~Bakas} {The Large $N$ Limit   of Extended
Conformal Symmetries}{\PLB{228}{1989}{57}.}{}}

\lref\wwwgravity{\paper {C.M.~Hull}{Lectures on $\CW$-Gravity,
$\CW$-Geometry and
$\CW$-Strings}{}{\hh 9302110}, and~references therein.}

\lref\wvgravity{\paper {A.~Bilal, V.~Fock, I.~Kogan}{On the origin of
$W$-algebras}{\NPB{359}{1991}{635}.}{}}

\lref\surprises{\paper {E.~Witten} {Surprises with Topological Field
Theories} {Lectures given at ``Strings 90'', Texas A\&M, 1990,}{Preprint
IASSNS-HEP-90/37.}}

\lref\stringsone{\paper {L.~Baulieu, M.B.~Green, E.~Rabinovici}{A Unifying
Topological Action for Heterotic and  Type II Superstring  Theories}
{\PLB{386}{1996}{91},}{\hh 9606080.}}

\lref\stringsN{\paper {L.~Baulieu, M.B.~Green, E.~Rabinovici}{Superstrings
from   Theories with $N>1$ World Sheet Supersymmetry}
{\NPB{498}{1997}{119},}{\hh 9611136.}}

\lref\bks{\paper {L.~Baulieu, H.~Kanno, I.~Singer}{Special Quantum Field
Theories in Eight and Other Dimensions}{\CMP{194}{1998}{149},}{\hh
9704167}\semi
\paper {L.~Baulieu, H.~Kanno, I.~Singer}{Cohomological Yang--Mills Theory
  in Eight Dimensions}{ Talk given at APCTP Winter School on Dualities in
String Theory (Sokcho, Korea, February 24-28, 1997),} {\hh 9705127.}}

\lref\witdyn{\paper {P.~Townsend}{The eleven dimensional supermembrane
revisited}{\PLB{350}{1995}{184},}{\hh9501068}\semi
\paper{E.~Witten}{String Theory Dynamics in Various Dimensions}
{\NPB{443}{1995}{85},}{\hh 9503124}.}

\lref\bfss{\paper {T.~Banks, W.Fischler, S.H.~Shenker,
L.~Susskind}{$M$-Theory as a Matrix Model~:
A~Conjecture}{\PRD{55}{1997}{5112},} {\hh9610043.}}

\lref\seiberg{\paper {N.~Seiberg}{Why is the Matrix Model
Correct?}{\PRL{79}{1997}{3577},} {\hh 9710009.}}

\lref\sen{\paper {A.~Sen}{$D0$ Branes on $T^n$ and Matrix Theory}{Adv.
Theor. Math. Phys.~{\bf 2} (1998) 51,} {\hh 9709220.}}

\lref\laroche{\paper {L.~Baulieu, C.~Laroche} {On Generalized Self-Duality
Equations Towards Supersymmetric   Quantum Field Theories Of
Forms}{\MPL{A13}{1998}{1115},}{\hh  9801014.}}

\lref\bsv{\paper {M.~Bershadsky, V.~Sadov, C.~Vafa} {$D$-Branes and
Topological Field Theories}{\NPB{463} {1996}{420},}{\hh9511222.}}

\lref\vafapuzz{\paper {C.~Vafa}{Puzzles at Large N}{}{\hh 9804172.}}

\lref\dvv{\paper {R.~Dijkgraaf, E.~Verlinde, H.~Verlinde} {Matrix String
Theory}{\NPB{500}{1997}{43},} {\hh9703030.}}

\lref\wynter{\paper {T.~Wynter}{Gauge Fields and Interactions in Matrix
String Theory}{\PLB{415}{1997}{349},}{\hh9709029.}}

\lref\kvh{\paper {I.~Kostov, P.~Vanhove}{Matrix String Partition
Functions}{}{\hh9809130.}}

\lref\ikkt{\paper {N.~Ishibashi, H.~Kawai, Y.~Kitazawa, A.~Tsuchiya} {A
Large $N$ Reduced Model as Superstring}{\NPB{498} {1997}{467},}{\hh
9612115.}}

\lref\ss{\paper {S.~Sethi, M.~Stern} {$D$-Brane Bound States
Redux}{\CMP{194}{1998} {675},}{\hh 9705046.}}

\lref\mns{\paper {G.~Moore, N.~Nekrasov, S.~Shatashvili} {$D$-particle Bound
States and Generalized Instantons}{} {\hh 9803265.}}

\lref\bsh{\paper {L.~Baulieu, S.~Shatashvili} {Duality from Topological
Symmetry}{} {\hh 9811198.}}

\lref\pawu{ {G.~Parisi, Y.S.~Wu} {}{ Sci. Sinica  {\bf 24} {(1981)} {484}.}}

\lref\coulomb{ {L.~Baulieu, D.~Zwanziger, }   {\it Renormalizable Non-Covariant
Gauges and Coulomb Gauge Limit}, {Nucl.Phys. B {\bf 548 } (1999) 527,} {\hh
9807024}.}

\lref\szczoth{ {A. Szczepaniak,} {arXiv: hep-ph/0306030}. }

\lref\dan{ {D.~Zwanziger},  {}{Nucl. Phys. B {\bf   139}, (1978) {1}.}{}}

\lref\danzinn{  {J.~Zinn-Justin, D.~Zwanziger, } {}{Nucl. Phys. B  {\bf
295} (1988) {297}.}{}}

\lref\danlau{ {L.~Baulieu, D.~Zwanziger, } {\it Equivalence of Stochastic
Quantization and the Faddeev-Popov Ansatz,
  }{Nucl. Phys. B  {\bf 193 } (1981) {163}.}{}}

\lref\munoz{ { A.~Munoz Sudupe, R. F. Alvarez-Estrada, } {}
Phys. Lett. {\bf 164} (1985) 102; {} {\bf 166B} (1986) 186. }

\lref\okano{ { K.~Okano, } {}
Nucl. Phys. {\bf B289} (1987) 109; {} Prog. Theor. Phys.
suppl. {\bf 111} (1993) 203. }

\lref\singer{
 I.M. Singer, { Comm. of Math. Phys. {\bf 60} (1978) 7.}}

\lref\neu{ {H.~Neuberger,} {Phys. Lett. B {\bf 295}
(1987) {337}.}{}}

\lref\testa{ {M.~Testa,} {}{Phys. Lett. B {\bf 429}
(1998) {349}.}{}}

\lref\Martin{ L.~Baulieu and M. Schaden, {\it Gauge Group TQFT and Improved
Perturbative Yang--Mills Theory}, {  Int. Jour. Mod.  Phys. A {\bf  13}
(1998) 985},   hep-th/9601039.}

\lref\baugros{ {L.~Baulieu, B.~Grossman, } {\it A topological Interpretation
of  Stochastic Quantization} {Phys. Lett. B {\bf  212} {(1988)} {351}.}}

\lref\bautop{ {L.~Baulieu}{ \it Stochastic and Topological Field Theories},
{Phys. Lett. B {\bf   232} (1989) {479}}{}; {}{ \it Topological Field Theories
And Gauge Invariance in Stochastic Quantization}, {Int. Jour. Mod.  Phys. A
{\bf  6} (1991) {2793}.}{}}

\lref\samson{ {L.~Baulieu, S.L.~Shatashvili, { \it Duality from Topological
Symmetry}, {JHEP {\bf 9903} (1999) 011, hep-th/9811198.}}}{}

\lref\halpern{ {H.S.~Chan, M.B.~Halpern}{}, {Phys. Rev. D {\bf   33} (1986)
{540}.}}

\lref\yue{ {Yue-Yu}, {Phys. Rev. D {\bf   33} (1989) {540}.}}

\lref\neuberger{ {H.~Neuberger,} {\it Non-perturbative gauge Invariance},
{ Phys. Lett. B {\bf 175} (1986) {69}.}{}}

\lref\schwinger{  {J.~Schwinger,} {}{Phys. Rev. {\bf 127} (1962)
{324}.}{}}

\lref\gribov{  {V.N.~Gribov,} {}{Nucl. Phys. B {\bf 139} (1978) {1}.}{}}

\lref\huffel{ {P.H.~Daamgard, H. Huffel},  {}{Phys. Rep. {\bf 152} (1987)
{227}.}{}}

\lref\creutz{ {M.~Creutz},  {\it Quarks, Gluons and  Lattices, }  Cambridge
University Press 1983, pp 101-107.}

\lref\zinn{ {J. ~Zinn-Justin, }  {Nucl. Phys. B {\bf  275} (1986) {135}.}}

\lref\shamir{  {Y.~Shamir,  } {\it Lattice Chiral Fermions
  }{ Nucl.  Phys.  Proc.  Suppl.  {\bf } 47 (1996) 212,  hep-lat/9509023;
V.~Furman, Y.~Shamir, Nucl.Phys. B {\bf 439 } (1995), hep-lat/9405004.}}

 \lref\kaplan{ {D.B.~Kaplan, }  {\it A Method for Simulating Chiral
Fermions on the Lattice,} Phys. Lett. B {\bf 288} (1992) 342; {\it Chiral
Fermions on the Lattice,}  {  Nucl. Phys. B, Proc. Suppl.  {\bf 30} (1993)
597.}}

\lref\neubergerr{ {H.~Neuberger, } {\it Chirality on the Lattice},
hep-lat/9808036.}

\lref\zbgr {L.~Baulieu and D. Zwanziger, {\it QCD $_4$ From a
Five-Dimensional Point of View},    hep-th/9909006.}

\lref\christlee{ {N. ~Christ and T. ~D. ~Lee,} {\it } {Phys. Rev.
{\bf  D22} (1980) {939}.}}

\lref\lattcoul{ {D. ~Zwanziger,} {\it Lattice Coulomb hamiltonian and
static color-Coulomb field,} {Nucl. Phys. B {\bf  485} (1997) {185}.}}

\lref\cnltcoul{ {D. ~Zwanziger,} {\it Continuum and Lattice
Coulomb-gauge hamiltonian,} {In *Cambridge 1997, Confinement, duality,
and non-perturbative aspects of QCD*, P. van Baal, Ed.}
{hep-th/9710157.}}

\lref\coul{ {D. ~Zwanziger,} 
{Nucl. Phys. B {\bf  518} (1998) {237}.}}

\lref\recoul{ {L.~Baulieu and D. ~Zwanziger,} {\it Renormalizable non-covariant
gauges and Coulomb gauge limit,} {Nucl. Phys. B {\bf  548} (1999) {527}.}}

\lref\critical{ {D. ~Zwanziger,} {\it Critical Limit of Lattice Gauge Theory,} {Nucl.
Phys. B {\bf  378} (1992) {525}.}}

\lref\vanish{ {D. ~Zwanziger,} {\it Vanishing of zero-momentum lattice gluon
propagator and color confinement,} {Nucl. Phys. B {\bf  364} (1991) {127}.}}

\lref\DA{ {A. Cucchieri and D. Zwanziger,} {Nucl. Phys. Proc. Suppl.
119 (2003) 727, arXiv: hep-lat/0209068.}}

\lref\noconfw{ {D. Zwanziger,}{ Phys. Rev. Lett. 90 (2003) 102001;
arXiv: hep-lat/0209105.} }

\lref\dznonpland{ {D.~Zwanziger,}   
Phys. Rev. D, {\bf 65} 094039 (2002) and
hep-th/0109224.}

\lref\npertfp{ {D. Zwanziger,}{ Phys. Rev. D (to be published);
arXiv: hep-ph/0303028.} }

\lref\JS{ {J. Greensite and {\v S}. Olejn\'{\i}k,} {Phys. Rev. D67
(2003) 094503; arXiv: hep-lat/0302018, and arXiv: hep-lat/0309172.} }

\lref\smekal{L.~von Smekal, A.~Hauck and R.~Alkofer,  
Ann. Phys. {\bf 267} (1998) 1; 
L. von Smekal, A. Hauck and R. Alkofer, 
Phys. Rev. Lett. {\bf 79} (1997) 3591; 
L. von Smekal 
Habilitationsschrift, Friedrich-Alexander
Universit\"{a}t, Erlangen-N\"{u}rnberg (1998).}

\lref\smekrev {R.~Alkofer and L.~von Smekal, 
Phys. Rept. {\bf 353}, 281 (2001).}

\lref\fischalk {C. S. Fischer and R.~Alkofer, {\it Infrared exponents and
running coupling of SU(N) Yang-Mills Theories},   Phys. Lett. B {\bf 536},
177 (2002).}

\lref\fischalkrein{C.~S.~Fischer, R.~Alkofer and H.~Reinhardt,
   {\it The elusiveness of infrared critical exponents in Landau gauge
   Yang-Mills theories,}
   Phys. Rev. D {\bf 65}, 094008 (2002)}

\lref\fischalkqu{C.~S.~Fischer and R.~Alkofer,
   {\it Non-perturbative propagators, running coupling and dynamical quark
mass of Landau gauge QCD,}
   hep-ph/0301094}

\lref\alkdetfischmar{R.~Alkofer, W.~Detmold, C.~S.~Fischer, P.~Maris,
   arXiv: hep-ph/0309077}

\lref\lerche {C. Lerche and L. von Smekal, 
arXiv: hep-ph/0202194}

\lref\Robertson{ {D. G. Robertson, E. S. Swanson, A. P. Szczepaniak,
C.-R. Ji, S. R. Cotanch, } {\it  Renormalized Effective QCD Hamiltonian:
Gluonic Sector, }{Phys. Rev. D59 (1999) 074019.}}	

\lref\Szcz{ {Adam Szczepaniak, Eric S. Swanson, Chueng-Ryong Ji, Stephen R.
Cotanch, } {\it  Glueball Spectroscopy in a Relativistic Many-Body Approach to
Hadron Structure, }{Phys. Rev. Lett. 76 (1996) 2011-2014.}}	

\lref\cuzwsc{ {Attilio Cucchieri, Daniel Zwanziger, } {\it  Static
Color-Coulomb Force, }{Phys. Rev. Lett. 78 (1997) 3814-3817.}}	

\lref\ZZ{ {Ismail Zahed, Daniel Zwanziger, } {\it  Zero Color Magnetization in
QCD Matter, }{Phys. Rev. D61 (2000) 037501.}}	

\lref\cuzwns{ {Attilio Cucchieri, Daniel Zwanziger, } 
{Phys. Rev. D65 (2001) 014001.}}	

\lref\rengrcoul{ {Attilio Cucchieri, Daniel Zwanziger, } 
{Phys. Rev. D65 (2001) 014002.}}	

\lref\pesschro{ {Michael Peskin, Daniel Schroeder, } {\it  An
Introduction to field theory, }{Perseus Books (1995) p. 593.}}	

\lref\doust{ {R. N. Doust, } {\it  Ann. of Phys., }{177 (1987) 169.}}	

\lref\taylor{ {J. C. Taylor, } {\it Physical and Non-standard Gauges,}
{Proc., Vienna, Austria, 1989, ed. P. Gaigg, W. Kummer and M.
Schweda (Springer, Berlin, 1990) p. 137.}}	

\lref\tdlee{ {T. D. Lee, } {\it  Particle physics and
introduction to field theory, }{Harwood (1981) p. 455.}}


 


\Title{\vbox
{\baselineskip 10pt
\hbox{hep-th/0312254}
\hbox{NYU-TH-PH-20.8.00}
\hbox{BI-TP 2000/19}
 \hbox{   }
}}
{\vbox{\vskip -30 true pt
\centerline{ Analytic calculation of color-Coulomb potential}
\vskip4pt }}
\centerline{
{\bf  Daniel Zwanziger}
}
\centerline{
daniel.zwanziger@nyu.edu}
\vskip 0.5cm

\centerline{\it 
Physics Department, New York University, New-York,  NY 10003,  USA}

\medskip
\vskip  1cm
\noindent

	We develop a calculational scheme in Coulomb and
temporal gauge that respects gauge invariance and is most easily
applied to the infrared asymptotic region of QCD.
It resembles the Dyson-Schwinger
equations of Euclidean quantum field theory in Landau gauge, but is
3-dimensional.  A simple calculation yields a 
color-Coulomb potential that behaves at large $R$ approximately like 
$V_{\rm coul}(R) \sim R^{[1-0.2(d-1)]}$ for spatial dimension 
$1 \leq d \leq 3$.  This   is a linearly rising potential plus a 
rather weak dependence on $d$.

\Date{\ }

\def\e{\epsilon}
\def\demi{{1\over 2}}
\def\quart{{1\over 4}}

\def\a{\alpha}

\def\d{\delta}

\def\om{\omega}

\def\s{\sigma}
\def\l{\lambda}
\def\L{\Lambda}

\newsec{Introduction}

 There is a simple confinement scenario in Coulomb gauge \gribov,
\coul\ which, in short, attributes confinement to the
long range of the color-Coulomb potential,~$V_{\rm coul}(R)$.  This
quantity is the instantaneous part of the 00-component of the dressed
gluon propagator in minimal Coulomb gauge,\foot{In this equation
$\vec{x}$ represents a 3-vector, but everywhere else in this article
3-vectors are represented by $x$.}
\eqn\doo{\eqalign{
g_0^2D_{00}(\vec{x}, x_0) = 
\langle gA_0^a(\vec{x},x_0) gA_0^b(0,0) \rangle
= V_{\rm coul}(|\vec{x} |) \d(x_0 ) +
({\rm non-instantaneous}),  
}}
and is given by \rengrcoul\
\eqn\potential{\eqalign{
V_{\rm coul}(x-y) \d^{ab} 
= \langle \ g_0^2 [M^{-1}(A)(-\p^2)M^{-1}(A)]_{xy}^{ab} \
\rangle.  }}
Here~$M(A) \equiv - \p_i D_i(A)$ is the Faddeev-Popov
operator, and the gauge-covariant derivative is defined by
$[D_i(A)\om]^a = \p_i \om^a + g_0f^{abc}A_i^b\om^c$.

We present a calculation of $V_{\rm coul}(R)$.
This quantity is of interest because:  
(i)~It couples universally to color charge.
(ii)~Confinement of color-charge may be explained by the long range
of this potential. 
(iii)~It is a renormalization-group
invariant and is independent of the cut-off and the renormalization
mass \coul.  
 (iv)~A necessary condition for the Wilson potential $V(R)$ to be
confining is that $V_{\rm coul}(R)$ be confining~\noconfw,
and if both potentials rise linearly at large $R$,
$V(R) \sim \s R$ and $V_{\rm coul}(R) \sim \s_{\rm coul}R$,
then $\s_{\rm coul} \geq \s$.
(v)~We wish to compare with a recent numerical
determination 
\JS\ of $V_{\rm coul}(R)$, that does show a linear rise at
large~$R$, with $\s_{\rm coul} \sim 3 \s$. 

Calculations in the Coulomb gauge have been pursued vigorously.  For
recent work and further references, see \szczoth.  The present approach
is distinguished by particular attention to gauge invariance, and its
easiest application is to the infrared asymptotic limit of QCD.

\newsec{Temporal gauge and Coulomb gauge}

For simplicity we consider pure gluodynamics.  
In the temporal or Weyl gauge, $A_0 = 0$, 
the wave functionals $\Psi(A)$
depend on $A_i^a(x)$ for
$i =1,2,3$.  The color-electric field operator is represented by 
$E_i^a(x) = i \d / \d A_i^a(x)$, and the hamiltonian by
\eqn\hamiltonian{\eqalign{
H \equiv \demi\int d^3x \ (E^2 + B^2),
}}
where $B_i^a = \e_{ijk}(\p_j A_k^a + \demi g_0 f^{abc} A_j^b A_k^c)$,
and the $f^{abc}$ are the structure constants of the Lie algebra of
the SU(N) group.  Wave functionals in temporal gauge are required to
be  gauge invariant $\Psi({^g}A) = \Psi(A)$,
where $g(x) \in SU(N)$ is a 3-dimensional local gauge
transformation, and
${^g}A_i \equiv  g_0^{-1}g^{-1} \p_i g + g^{-1}A_i g$.
These continuum equations
have precise analogs in lattice gauge theory, where the Kogut-Suskind
hamiltonian replaces the Weyl hamiltonian.

Poincare invariance of the continuum theory
is preserved because the hamiltonian density
$T^{00} = \demi(E^2 + B^2)$ satisfies the Dirac-Schwinger
equal-time commutation relation
\eqn\diracschwing{\eqalign{
[T^{00}(x), T^{00}(y)] = -i [T^{0i}(x) + T^{0i}(y)]\p_i \d(x-y) + 
{\rm S.T.},
}}
where $T^{0i} = \demi\e_{ijk}(E_j^a B_k^a + B_k^a E_j^a)$ is the
Poynting vector \schwinger, and S.T. is the Schwinger term.

Inner products in temporal gauge,
$(\Psi_1, \Psi_2) = N \int dA \ \Psi_1^*(A)\Psi_2(A)$,
are divergent because of the gauge
invariance of the wave-functionals.  They may be made finite by using
the Faddeev-Popov identity 
\eqn\innera{\eqalign{
(\Psi_1, \Psi_2) = \int_\L dA^{\rm tr} \ \det M(A^{\rm tr}) \
\Psi_1^*(A^{\rm tr})\Psi_2(A^{\rm tr}).
}}
The integral extends over 3-dimensionally 
transverse configurations in the
fundamental modular region $\L$, which is a region free of Gribov
copies.  To be definite we suppose that we are in the minimal
Coulomb gauge, which is obtained by minimizing 
$F_A(g) = ||{^g}A||^2$ with respect to gauge transformations $g(x)$,
so $||A|| \leq ||{^g}A||$ for all $g(x)$ and all $A$ in~$\L$.   

Wave-functionals in minimal Coulomb gauge
$\Psi(A^{\rm tr})$ are the restriction of
gauge-invariant wave functionals in temporal gauge
$\Psi(A)$ to the fundamental modular region~$\L$.
Conversely every wave-functional in minimal Coulomb gauge has a
unique gauge-invariant extension to temporal gauge. 
Every point $A$ in the interior of $\L$, is a unique 
absolute minimum (modulo global gauge transformations), so the strict
inequality holds $||A|| < ||{^g}A||$ (for all $g(x)$ that is not a
global gauge transformation).  
But every point $A_1$ on the boundary $\p \L$ of $\L$ is related by
a local gauge transformation $g(x)$
to some other point $A_2 = {^g}A_1$ also on 
$\p \L$, with which it is degenerate, $||A_1|| = ||A_2||$.
This gauge transformation may be infinitesimal,
$A_2 = A_1 + \e D(A_1)\om$, where $D(A_1)\om$ is tangent to $\p \L$.
Gauge-invariance 
requires that the wave-functional in Coulomb gauge be identified at
corresponding boundary points, 
$\Psi(A_2) = \Psi(A_1)$, or, for the infinitesimal case, that the
wave functional satisfies
$(D(A_1)\om, { {\d \Psi} \over {\d A} }|_{A_1}) = 0$.  This provides
the boundary condition that is needed to make the hamiltonian in
Coulomb gauge well-defined and symmetric.  The identification of
boundary points is often ignored because one does not know explicitly
what the boundary of $\L$ is.  But in general it would be a violation
of gauge invariance to ignore this identification and
take arbitrary wave-functionals in so-called
physical coordinates which are the transverse configurations in~$\L$.  
In order not to make this error, in the present
article we shall use wave functionals $\Psi(A)$ that are manifestly
gauge-invariant.  

As an example we exhibit an approximate vacuum wave functional that is
gauge invariant.  The variation of
the color-magnetic field is given by
\eqn\differentiate{\eqalign{
\d B_i^a = \e_{ijk}D_j^{ac} \d A_k^c \equiv (\hat{D} \d A)_i^a,
}}
which defines the hermitian operator $\hat{D}(A)$ that is the
gauge-covariant curl.  Consider
the wave functional
\eqn\wave{\eqalign{
\Phi = \exp\Big(-\demi \int d^3x 
\ B_i^a \ [(\hat{D}^2)^{-1/2} B]_i^a \Big).
}}
The operator $\hat{D}(A)$ has small eigenvalues when acting on
longitudinal fields, but the Bianchi identity $D_iB_i = 0$ insures
that the wave-functional is regular. 
 We have 
${ {\d \Phi} \over {\d A_i} } \approx
 - [\hat{D}(\hat{D}^2)^{-1/2} B]_i\Phi$,
and 
\eqn\schrod{\eqalign{
-\demi\int d^3x \ { {\d^2 \Phi} \over {\d A_i^2} } 
\approx \int d^3x \ (- \demi B^2 + f ) \ \Phi,
}}
where  
$f(x) \equiv \demi 
[(\hat{D}^2)^{1/2}]_{ii}^{aa}(x,y)|_{y=x}$,
and $\approx$ means that derivatives with respect to
$(\hat{D}^2)^{-1/2}$ are neglected.  The first term will cancel the
magnetic energy density, which is the most singular term,
being the product of quantum fields at the same point.
We have $f(x) = e + u(x, A)$, where 
$e = \demi [(\hat{\p}^2)^{1/2}]_{ii}^{aa}(x,y)|_{y=x}$ 
is a divergent constant,
and $u(x,A)$ is a gauge-invariant non-local functional that vanishes
with $A$.  The Schr\"{o}dinger equation reads
\eqn\differentiate{\eqalign{
H \Phi \approx \int d^3x \ [e + u(x,A)] \Phi,
}}
and is violated by non-local terms only.

\newsec{Calculational scheme}

	The vacuum wave functional $\Psi_0(A)$ is positive,
and we write $\Psi_0(A) = \exp[-S(A)/2]$,
where $S(A)$ is manifestly gauge invariant.  
We assume that $S(A)$ is either an approximate expression, such as the
one given above, or a trial expression.  With gauge-invariant
wave-functionals it is difficult to evaluate matrix elements
by direct integration, and we shall borrow techniques from
Euclidean quantum field theory.  We have
$|\Psi_0(A)|^2 = \exp[-S(A)]$,
and we define the generating functional of equal-time correlators,
\eqn\partition{\eqalign{
Z(J) \equiv \int_\L dA^{\rm tr} \exp(J, A^{\rm tr}) 
\ \det M(A^{\rm tr}) \ \exp[-S(A^{\rm tr})] ,
}}
normalized to $Z(0) = 1$.  This is precisely the formula for the
partition function or generating functional of 3-dimensional Euclidean
gauge theory in the minimal Landau gauge, with the Yang-Mills Euclidean
action $S_{\rm YM}(A)$ replaced by some
gauge-invariant action $S(A)$.  Only the transverse part of the source 
$J_i^a(x)$ contributes, and we take $J_i$ to be identically
transverse, $\p_i J_i = 0$, and $J_i = J_i^{\rm tr}$. 

We don't have an explicit expression for~$\L$, and we
rely on the argument of~\npertfp\
that fundamental modular region $\L$ and the Gribov region
$\Omega$ have the same moments or correlators so we may
integrate over~$\Omega$ instead of~$\L$,  
\eqn\partitiona{\eqalign{
Z(J) \equiv \int_\Omega dA^{\rm tr} \exp(J, A^{\rm tr}) 
\ \det M(A^{\rm tr}) \ \exp[-S(A^{\rm tr})] .
}}
Whereas $\L$ is the set of absolute minima of the minimizing
functional, the Gribov region~$\Omega$
is the set of relative minima.
The matrix of second derivatives of the minimizing functional
is the Faddeev-Popov operator~$M(A)$. 
It is a non-negative matrix at a relative minimum, so
the Gribov region~$\Omega$
is the set of transverse configurations $A^{\rm tr}$ for which
all eigenvalues~$\l_n(A^{\rm tr})$ 
of $M(A^{\rm tr})$ are non-negative,
$\l_n(A^{\rm tr}) \geq 0$.  The interior of $\Omega$
consists of points $A^{\rm tr}$ where all eigenvalues
are strictly positive, $\l_n(A^{\rm tr}) > 0$ (apart from a trivial
null eigenvector corresponding to global gauge transformations).
Its boundary $\p \Omega$ consists of points where
$M(A^{\rm tr})$ has a non-trivial null-eigenvector,
$M(A^{\rm tr}) \om = 0$, 
so $\l_1(A^{\rm tr}) = 0$, and
all other eigenvalues are non-negative, $\l_n(A^{\rm tr}) \geq 0$, for 
$A^{\rm tr} \in \p \Omega$.   

	We don't have an explicit expression for $\Omega$ either,
but we may exploit the fact that the integrand of~\partitiona\
vanishes on~$\p \Omega$ 
to derive the Dyson-Schwinger (DS) equations nevertheless \dznonpland.
Indeed the Faddeev-Popov determinent vanishes 
for $A^{\rm tr} \in \p \Omega$,
$\det M(A^{\rm tr}) = \prod_n\l_n(A^{\rm tr}) = 0$.  
Thus the identity 
\eqn\identity{\eqalign{
0 = \int_\Omega dA^{\rm tr} 
{ { \d  } \over { \d A^{\rm tr}_i(x) } } 
\Big( \exp(J, A^{\rm tr}) 
\ \det M(A^{\rm tr}) \ \exp[-S(A^{\rm tr})] \Big),
}}
holds without a contribution from the boundary $\p \Omega$.  The
set of DS equations
in functional form,
\eqn\dsz{\eqalign{
{ {\d \Sigma } \over { \d A_i^{{\rm tr},a}(x) } }
\Big({ \d  \over {\d J} } \Big) \ Z(J) = J_i^a(x) \ Z(J),
}} 
follow from this identity.  Here 
$\Sigma(A^{\rm tr}) \equiv S(A^{\rm tr}) - {\rm tr} \ln M(A^{\rm tr})$ 
is the effective action.  Because the integrand vanishes on 
$\p \Omega$, the DS equations have the same form as they
would if the integral~\identity\ were extended to 
infinity.  It is not necessary to know 
the boundary~$\p \Omega$ explicitly, and the cut-off 
at~$\p \Omega$ is implemented by imposing 
on the solution of the DS equations the natural positivity
conditions that must be satisfied by the (equal-time) correlator
$\langle A_i(x) A_j(y) \rangle$, by the ghost propagator
$G(x-y) = \langle g_0 M^{-1}(A^{\rm tr})(x,y) \rangle$, 
and by the higher-order
correlators.

	As in Euclidean quantum field theory, 
we rewrite \dsz\ as a functional DS equation for 
$W(J) \equiv \ln Z(J)$, which is the analog of the free energy
\eqn\dsw{\eqalign{
{ {\d \Sigma } \over { \d A_i^{{\rm tr},a}(x) } }
\Big({ \d W  \over {\d J} } + { \d  \over {\d J} } \Big) \ 1
 = J_i^a(x).
}} 
By Legendre transformation we convert this to a functional DS
equation for the analog of the quantum effective action
$\G(A^{\rm tr}) \equiv (A^{\rm tr}, J) - W(J)$,
where 
$A_i^{{\rm tr},a}(x) \equiv 
{ {\d W(J)} \over { \d J_i^a(x)} }$,
\eqn\dsgamma{\eqalign{
{ {\d \Sigma } \over { \d A_i^{{\rm tr},a}(x) } }
\Big(A^{\rm tr} + {\cal D}{ \d  \over {\d A^{\rm tr}} } \Big) \ 1
 = { {\d \G } \over { \d A_i^{{\rm tr},a}(x) } },
}} 
where ${\cal D}$ is the gluon propagator in the presence of the
source, 
${\cal D}^{-1}(A^{\rm tr}) 
= { { \d^2 \G} \over { \d A^{\rm tr} \d A^{\rm tr}} }$.
The problem of evaluating correlators by
direct functional integration has been replaced by the problem of
solving the DS equations and, given the action $S(A)$,
we can, at least in principle, calculate
all correlators by solving the DS
equations for $\G(A^{\rm tr})$.

Suppose we take a trial expression $S(A, \xi)$ for the gauge-invariant
action that depends on some unknown parameters $\xi$.
	These parameters are determined by minimizing
$E(\xi) \equiv \langle H \rangle =  (\Psi_0, H \Psi_0)$.  
To calculate~$E(\xi)$, write $H = H_{\rm e} + H_{\rm m}$.  We have
$H_{\rm m} = \demi\int d^3x \ B^2(x)$, where 
$B_i^a(x) = B_i^a(x; A^{\rm tr})$, and the magnetic energy is given by
\eqn\magenergy{\eqalign{
E_{\rm m}(\xi) = \langle H_{\rm m} \rangle = \demi \int d^3x \ 
B^2\Big(x; { \d \over {\d J} } \Big) \ Z(J)|_{J=0}.
}}
To calculate the electric energy 
\eqn\elecenergy{\eqalign{
E_{\rm e}(\xi) = \langle H_{\rm e} \rangle  = \demi \int d^3x 
\int_\Omega dA^{\rm tr} \ \det M(A^{\rm tr}) \
  |E(x)\Psi_0|^2,  
}}
we evaluate $E_i\Psi$ in temporal gauge, and then restrict
to $\Omega$.  With $\Psi_0(A) = \exp[-\demi S(A)] $, we have
$E_i^a(x)\Psi_0 = i { {\d \Psi_0} \over { \d A_i^a(x)} } =
- i {\cal E}_i^a(x; A) \Psi_0,$
where
${\cal E}_i^a(x; A) \equiv \demi { {\d S(A)} \over { \d A_i^a(x) } }$, 
which gives
\eqn\elecenergya{\eqalign{
E_{\rm e}(\xi) & = \demi \int d^3x 
\int_\L dA^{\rm tr} \ \det M(A^{\rm tr}) \
  {\cal E}^2(x; A^{\rm tr})
\exp[-S(A^{\rm tr})] \cr
& = \demi \int d^3x \
  {\cal E}^2\Big(x; { \d \over {\d J} } \Big) \ Z(J)|_{J=0}. 
}}
Now $E(\xi) = E_{\rm e}(\xi) + E_{\rm m}(\xi)$ has, in principle, been
expressed in terms of the
$\xi$ parameters that appear in $S(A, \xi)$.  

\newsec{Infrared Ansatz}

	This program may be difficult to carry out, especially if the 
action $S(A)$ is non-local.  However if our experience with 
DS equations with action $S_{\rm YM}(A)$ is a reliable guide, then a
remarkable simplification occurs in the infrared limit, as we now
explain.  

The DS equations with Yang-Mills action were first solved, 
with due attention to the ghost contribution, in~\smekal.  The
subject is reviewed in \smekrev, and recent results are reported
in~\alkdetfischmar.  We will follow the method
of~\lerche,~\dznonpland.  It was found in these investigations that
the ghost contribution is the dominant one in the infrared.  For
example, in the DS equation for the gluon propagator, the  leading
contribution in the infrared is provided by the gluon loop. It was
subsequently realized~\npertfp\ that in the DS equation 
\dsgamma, with effective action
$\Sigma = S_{\rm YM} - {\rm tr} \ln M$, one obtains the correct
infrared asymptotic limit by setting $S_{\rm YM} = 0$. 
Thus the infrared asymptotic limit is entirely determined by the
Faddeev-Popov determinent $\det M(A^{\rm tr})$ and the cut-off at the
Gribov horizon $\p \Omega$.  One might think that the functional
integral with $S_{\rm YM} = 0$ would diverge.  However the DS
equations are merely a technique for evaluating the functional
integral, and since they give a finite result with $S_{\rm YM} = 0$,
it appears that cut-off at the Gribov horizon makes the functional
integral converge.

{\it Infrared Ansatz:} We shall assume that in the present case
also, the correct infrared limit is obtained by setting 
$S(A) = 0$. Moreover once one sets $S(A) = 0$,
the present calculation reduces
to the calculation of the infrared limit in Landau gauge, 
where one has $S_{\rm YM}(A) = 0$, 
and we may use directly the solution of \dznonpland,
where $d$ now represents the dimension of space instead of the
dimension of space-time.

We briefly outline how the solution was
obtained in \dznonpland.  
The crucial point is that a solution was sought for which the ghost
propagator, $\tilde{G}(k)$, is more singular than
$1/k^2$ at $k=0$.  This property has been called the ``horizon
condition", and it triggers the confinement scenario in Coulomb
gauge.   The horizon condition holds because the Gribov
region~$\Omega$ is bounded in every direction and, in a space of high
dimension such as configuration space, entropy favors population
density near the Gribov horizon~$\p \Omega$.  At the horizon
$M^{-1}(A^{\rm tr})$ is singular, and this enhances the ghost
propagator
$G(x-y) = \langle g_0(M^{-1})_{xy}) \rangle $ at large separation
or small $k$.  The coupled DS equations
for the gluon and ghost propators were solved
in~\lerche, and~\dznonpland, taking the tree-level
expression for the ghost-gluon vertex, and imposing the transversality
condition $\p_i A_i = 0$ on-shell.

\newsec{Calculation of color-Coulomb potential}

To calculate $V_{\rm coul}(R)$ we write \potential\ as
\eqn\matrixprod{\eqalign{
V_{\rm coul}(x-y) \d^{ab}
= \int d^3z \ \langle \ {\cal G}^{ac}(x,z; A^{\rm tr})
\ (-\p_z^2) \ {\cal G}^{cb}(z,y; A^{\rm tr}) \ \rangle,
}}
and
${\cal G}^{ab}(x,y; A^{\rm tr}) 
\equiv g_0 [M^{-1}(A^{\rm tr})]_{xy}^{ab}$. 
Its expectation-value,
$G(x-y) \d^{ab} = \langle {\cal G}^{ab}(x,y; A^{\rm tr}) \rangle$,
is the ghost propagator. 
We separate the expectation-value of the product in \matrixprod\
into disconnected and connected parts, 
\eqn\potentiala{\eqalign{
V_{\rm coul}(x-y)   = 
\int d^3z \ G(x-z) \ (-\p_z^2) \ G(z-y)  + 
V_{\rm con}(x-y), 
}}
which reads, upon fourier transformation,
\eqn\momentum{\eqalign{
\tilde{V}_{\rm coul}(k) = 
k^2\tilde{G}^2(k) + \tilde{V}_{\rm con}(k). 
}}

	The infrared
asymptotic form of the gluon and ghost propagators
depends on two infrared critical exponents,
\eqn\propagators{\eqalign{
D^{\rm as}(k^2) & = { {b_D} \over  {(k^2)^{1 + \a_D} } } \cr
G^{\rm as}(k^2) & = { {b_G} \over  {(k^2)^{1 + \a_G} } } \ .
}}
By equating like powers of momentum in either the gluon or the
ghost DS equation, one obtains in either case the same relation
$\a_D + 2\a_G = - (4-d)/2$.  We use this equality to eliminate $\a_D$,
in favor of $\a \equiv \a_G$.  
In the infrared limit, only the ghost loop contributes to the
DS equation for the gluon which reads
gives $(b_D b_G^2)^{-1} = I_D(\a, d)$, where by eq.~(A6) 
of~\dznonpland
\eqn\eyedee{\eqalign{
I_D(\a, d) = { {N} \over {2 \ (4\pi)^{d/2} } }  
\ { {\G(2\a + 1 - d/2) \ \G^2(-\a + d/2)}
 \over {\G^2(1 + \a)\G(d - 2\a)} }.
}} 
The only process that contributes to the DS equation for the ghost
propagator is emission and absorption of a gluon.  In the infrared
limit this gives
$(b_D b_G^2)^{-1} = I_G(\a, d)$, where by eq.~(A17) of~\dznonpland
\eqn\eyegee{\eqalign{
I_G(\a, d) =   
     \ { {N \ (d-1) } 
\over { 2 \ (4 \pi)^{d/2} } }
  { {\pi} \over {\sin(\pi \a)} }    
   \ { {\G(2\a + 1) \ \G(- \a + d/2)} 
  \over {\G^2(\a + 1) \ \G(- 2\a + d/2) \ \G(\a + 1 + d/2)} }.
}}
We must solve 
\eqn\eqforalpha{\eqalign{
I_D(\a, d) = I_G(\a, d)
}} 
to find the infrared critical exponent $\a = \a(d)$.

\newsec{Discussion of solution}

	We are interested in spatial dimension $1 < d \leq 3$.  The integral
for $I_D(\a, d)$ converges for $\a$ in the interval 
$\quart (d-2) < \a < \demi d$, 
and $I_D(\a, d)$ is positive in this interval and diverges at the
end-points.   The integral for $I_G(\a, d)$ converges
for $\a$ in the interval $ 0 < \a < 1$, and it diverges at the
end-points.  However $I_G(\a, d)$ changes sign in this interval at
$\a = \quart d$, and is positive only for 
$0 < \a < \quart d$.  Thus we look for solutions
for $\a$ in the range
${\rm max}[0, \quart(d-2)] \leq \a \leq \quart d$.

First take $d$ in the interval $1 < d < 2$.  We have
$\quart (d-2) < 0$ so we restrict our consideration to
the interval $0 < \a < \quart d$.  From the values
at the end-points it follows that there are an odd number
of solutions, and from numerical plots one sees that there is precisely
one solution $\a(d)$ for $1 < d < 2$.   At $d = 1$, $I_G(\a, d)$
vanishes because of the coefficient $(d-1)$ in \eyegee.  This
coefficient occurs because the ghost emits and aborbs a gluon
whose propagator contains a~$d$-dimensional transverse projector,
which vanishes at $d=1$. (The coefficient $d-1$ is an exact property of
the DS equation of the ghost, and holds also in a more refined
evaluation.)   In contrast, $I_D(\a, d)$ is finite at 
$d = 1$ because a coefficient
$d - 1$ has been factored out of both sides of the 
DS equation for the gluon propagator.  So $\a(d)$ vanishes
linearly with $(d-1)$, and from \eqforalpha\ one obtains in the 
limit~$d \to 1$,
\eqn\deenearone{\eqalign{
\a(d) \to { 2 \over {\pi^2} }(d-1) \approx 0.20264 \
(d-1). }}
In fact this formula fits a solution $\a(d)$ of
\eqforalpha\ to about 2~\% accuracy in the entire interval
$1 \leq d \leq 4$.  At $d = 2$ the exact solution is given by
\eqn\deeistwo{\eqalign{
\a(2) = { 1 \over 5},
}}
which differs from \deenearone\ by 1.3~\%.  The fitting formula,
\eqn\simpleform{\eqalign{
\a_{f1}(d) = { 1 \over 5}(d-1),
}}
represents a solution $\a_1(d)$
to \eqforalpha\ with about 1~\% accuracy in the
interval $1 \leq d \leq 4$.

Now consider spatial dimension $2 \leq d \leq 3$.  We have
$0 \leq \quart(d-2)$, so we seek a solution in the interval 
$\quart(d-2) < \a < \quart d$.  From the values at the end-points,
there are now an even number of solutions, and from numerical plots it
appears that there are precisely two distinct real roots 
$\a_1(d)$ and $\a_2(d)$ for 
$2 < d \leq 3$, except possibly at $d = d_c \approx 2.662$ where there
appears to be one root at $\a(2.662) \approx 0.33095$, which is
thus a crossing point of the two roots.  
At the crossing point \simpleform\ gives
$\a_{f1}(2.662) = 0.33240$, which is
accurate to 0.5~\%.  At $d = 2$
the two roots are given by $\a_1(2) = {1\over5}$ and $\a_2(2) = 0$. 
Only the root $\a_1(2) = {1\over5}$ matches the one root in the
interval
$1 < d < 2$, so it is the physical one, and
the root $\a_1(2) = 0$ is spurious.  
Thus for $2 \leq d < d_c \approx 2.662$, the larger root is the
physical one.  For values $d > d_c$ above 
the crossing point we do not know which
of the two roots is physical.  At $d = 3$ the equality \eqforalpha\
simplifies to 
${ {32 \a(1-\a)[1 - \cot^2(\pi \a)]} \over {(3 + 2\a) (1+2\a)} } = 1$,
with roots  
\eqn\criticalexp{\eqalign{
\a_1(3) \approx 0.3976; \ \ \ \ \ \ \ \ \ \  \a_2(3) = \demi.
}}
The fitting formula \simpleform\ gives 
$\a_{f1}(3) = 0.4$, which agrees with the first root
to about 1~\%.  At $d = 4$, there are two roots, 
$\a_1(4) = { {93 - \sqrt{1201} }\over{98} } \approx 0.5953$, 
and $\a_2(4) = 1$. The fitting formula \simpleform\
gives $\a_{f1}(4) = 0.6$, still accurate to 1 \%.
The fitting formula for the second root 
\eqn\secondroot{\eqalign{
\a_{f2}(d) = {1 \over 2} (d-2)
}} 
is exact at $d$ = 2, 3, and 4.  The two fitting formulas cross
at $d = {8\over3} \approx 2.666$ and $\a = {1\over3} \approx 0.333$.

	From the relation $\a_D + 2\a_G = \demi(d-4)$, and the
fitting formula \simpleform,
we obtain the critical exponent of the gluon propagator,
$\a_D(d) \approx - {3\over2} + {1\over{10}}(d-1)$.  This has a rather
weak dependence on the spatial dimension $d$ and gives a gluon
propagator $D(k)$ that vanishes at $k = 0$
for $1 \leq d \leq 3$.  Thus the would-be transverse
physical gluon does not appear in the spectrum.

The critical exponent of the color-Coulomb potential is defined by
\eqn\defineav{\eqalign{
\tilde{V}_{\rm coul}^{\rm as}(k) = { 1 \over {(k^2)^{1 + \a_V} } }, 
}}
	Suppose for simplicity that we neglect the connected term in
\momentum\ in the infrared asymptotic limit, leaving for another
occasion an evaluation of this term.  Then we have in
this limit
\eqn\momentuma{\eqalign{
\tilde{V}_{\rm coul}^{\rm as}(k)  = 
{ {b_G^2} \over {(k^2)^{1 + 2\a_G} }}   , 
}}
and we obtain for the infrared critical exponant of the color-Coulomb
potential $\a_V = 2 \a_G$.
The color-Coulomb
potential is given at large $R$ by
$V_{\rm coul}(R) \sim R^{2-d+2\a_V}$.
If one uses the simple fitting formula \simpleform\ for $\a_G$, one
gets
$\a_V = {2\over5}(d-1)$, and 
$V_{\rm coul}(R) \sim R^{[1-0.2(d-1)]}$, which is a linear potential 
plus a rather weak dependence on $d$ for 
$1 \leq d \leq 3$. For comparison we note that if instead 
$\a_G(d) = {1\over4} \ (d-1)$, which is not so different from our
solution, then one gets for the critical exponent of the gluon
$\a_D = - {3\over2}$, and of the color-Coulomb potential
$\a_V = \demi(d-1)$.  This gives an exactly
linear potential 
$V_{\rm coul}(R) \sim R$, asymptotically at large $R$.

The second
solution at $d = 3$, namely $\a_2(3) = \demi$, yields
$\tilde{V}_{\rm coul}^{\rm as}(k) \sim 1/k^4$, which gives an exactly a
linearly rising color-Coulomb potential.  However this success must
be regarded as partly accidental because our solution does not give
an exactly linear potential at $d = 2$, and a correct
calculation should work for both $d = 2$ and $d = 3$.

The deviation from a linear potential should not be regarded as a
failure of the approach because our truncation scheme requires making
an educated guess for the ghost-gluon vertex.  We have chosen the
tree-level vertex, but different choices give slightly different
critical exponents \smekal.  So in our approach there is an
inherent uncertainty in the critical exponent of the
color-Coulomb potential.  
Moreover a correction to \momentuma\ may result from a more accurate
evaluation of \momentum.
However, granted these limitations, our results
are at least qualitatively correct, and capture the essential features
of a confining theory.  We used the horizon condition which is the
qualitative requirement that the ghost propagator be enhanced in the
infrared compared to $1/k^2$.  The dynamics of the DS equation
then determine the infrared critical exponents of the ghost and gluon
propagators and of the color-Coulomb potential.  These are consistent
with the confinement scenario in Coulomb gauge, which requires
an infrared suppressed gluon propagator and a long-range color-Coulomb
potential.  They are also in at least qualitative 
agreement with numerical studies \cuzwns, \JS.  
The color-Coulomb potential we have obtained is
confining and not far from linear for $1 \leq d \leq 3$.

\vskip .5cm
{\centerline{\bf Acknowledgments}}

It is a pleasure to thank Christian Fischer,
Carlo Piccioni, and Reinhard Alkofer for valuable discussions.
I am grateful to Reinhard Alkofer for his hospitality at 
Universit\"{a}t T\"{u}bingen,
where part of this work was done.  This research was partially
supported by the National Science Foundation under grant PHY-0099393.

\vskip 2cm

\footatend\vfill\supereject\immediate\closeout\rfile\writestoppt
\baselineskip=14pt\centerline{{\bf References}}\bigskip{\frenchspacing%
\parindent=20pt\escapechar=` \input refs.tmp\vfill\eject}\nonfrenchspacing



\bye